\documentclass[%
 aip,
 jap,twocolumn,
 amsmath,amssymb,
 reprint,%
]{revtex4-1}
\usepackage{color}
\usepackage[
	pdffitwindow=true,
	colorlinks=true,
	frenchlinks=false,
        linkcolor=blue,
	anchorcolor=blue,
        citecolor=blue,
        filecolor=blue,
        urlcolor=blue,
        bookmarks=true,
        bookmarksopen=true,
	bookmarksnumbered=true, 
        bookmarksopenlevel=1,
        plainpages=false,
        pdfpagelabels=true,
	breaklinks
]{hyperref}
\usepackage[per-mode=symbol,separate-uncertainty]{siunitx}
\usepackage{graphicx}
\usepackage{dcolumn}
\usepackage{bm}

\begin{document}
\preprint{AIP/123-QED}

\title[]{Loss mechanisms in superconducting thin film microwave resonators}

\author{Jan Goetz}
\email[]{jan.goetz@wmi.badw.de}
\affiliation{Walther-Mei{\ss}ner-Institut, Bayerische Akademie der Wissenschaften, 85748 Garching, Germany }
\affiliation{Physik-Department, Technische Universit\"{a}t M\"{u}nchen, 85748 Garching, Germany}
\author{Frank Deppe}
\affiliation{Walther-Mei{\ss}ner-Institut, Bayerische Akademie der Wissenschaften, 85748 Garching, Germany }
\affiliation{Physik-Department, Technische Universit\"{a}t M\"{u}nchen, 85748 Garching, Germany}
\affiliation{Nanosystems Initiative Munich (NIM), Schellingstra{\ss}e 4, 80799 M\"{u}nchen, Germany}
\author{Max Haeberlein}
\affiliation{Walther-Mei{\ss}ner-Institut, Bayerische Akademie der Wissenschaften, 85748 Garching, Germany }
\affiliation{Physik-Department, Technische Universit\"{a}t M\"{u}nchen, 85748 Garching, Germany}
\author{Friedrich Wulschner}
\affiliation{Walther-Mei{\ss}ner-Institut, Bayerische Akademie der Wissenschaften, 85748 Garching, Germany }
\affiliation{Physik-Department, Technische Universit\"{a}t M\"{u}nchen, 85748 Garching, Germany}
\author{Christoph W. Zollitsch}
\affiliation{Walther-Mei{\ss}ner-Institut, Bayerische Akademie der Wissenschaften, 85748 Garching, Germany }
\affiliation{Physik-Department, Technische Universit\"{a}t M\"{u}nchen, 85748 Garching, Germany}
\author{Sebastian Meier}
\affiliation{Walther-Mei{\ss}ner-Institut, Bayerische Akademie der Wissenschaften, 85748 Garching, Germany }
\affiliation{Physik-Department, Technische Universit\"{a}t M\"{u}nchen, 85748 Garching, Germany}
\author{Michael Fischer}
\affiliation{Walther-Mei{\ss}ner-Institut, Bayerische Akademie der Wissenschaften, 85748 Garching, Germany }
\affiliation{Physik-Department, Technische Universit\"{a}t M\"{u}nchen, 85748 Garching, Germany}
\author{Peter Eder}
\affiliation{Walther-Mei{\ss}ner-Institut, Bayerische Akademie der Wissenschaften, 85748 Garching, Germany }
\affiliation{Physik-Department, Technische Universit\"{a}t M\"{u}nchen, 85748 Garching, Germany}
\affiliation{Nanosystems Initiative Munich (NIM), Schellingstra{\ss}e 4, 80799 M\"{u}nchen, Germany}
\author{Edwar Xie}
\affiliation{Walther-Mei{\ss}ner-Institut, Bayerische Akademie der Wissenschaften, 85748 Garching, Germany }
\affiliation{Physik-Department, Technische Universit\"{a}t M\"{u}nchen, 85748 Garching, Germany}
\affiliation{Nanosystems Initiative Munich (NIM), Schellingstra{\ss}e 4, 80799 M\"{u}nchen, Germany}
\author{Kirill G. Fedorov}
\affiliation{Walther-Mei{\ss}ner-Institut, Bayerische Akademie der Wissenschaften, 85748 Garching, Germany }
\affiliation{Physik-Department, Technische Universit\"{a}t M\"{u}nchen, 85748 Garching, Germany}
\author{Edwin P. Menzel}
\affiliation{Walther-Mei{\ss}ner-Institut, Bayerische Akademie der Wissenschaften, 85748 Garching, Germany }
\affiliation{Physik-Department, Technische Universit\"{a}t M\"{u}nchen, 85748 Garching, Germany}
\author{Achim Marx}
\affiliation{Walther-Mei{\ss}ner-Institut, Bayerische Akademie der Wissenschaften, 85748 Garching, Germany }
\author{Rudolf Gross}
\email[]{rudolf.gross@wmi.badw.de}
\affiliation{Walther-Mei{\ss}ner-Institut, Bayerische Akademie der Wissenschaften, 85748 Garching, Germany }
\affiliation{Physik-Department, Technische Universit\"{a}t M\"{u}nchen, 85748 Garching, Germany}
\affiliation{Nanosystems Initiative Munich (NIM), Schellingstra{\ss}e 4, 80799 M\"{u}nchen, Germany}

\date{prel. version\today}

\begin{abstract}
We present a systematic analysis of the internal losses of superconducting coplanar waveguide microwave resonators based on niobium thin films on silicon substrates. In particular, we investigate losses introduced by Nb/Al interfaces in the center conductor, which is important for experiments where Al based Josephson junctions are integrated into Nb based circuits. We find that these interfaces can be a strong source for two-level state (TLS) losses, when the interfaces are not positioned at current nodes of the resonator. In addition to TLS losses, for resonators including Al, quasiparticle losses become relevant above \SI{200}{\milli\kelvin}. Finally, we investigate how losses generated by eddy currents in conductive material on the backside of the substrate can be minimized by using thick enough substrates or metals with high conductivity on the substrate backside.
\end{abstract}

\maketitle

\section{Introduction}
Superconducting microwave transmission line resonators are widely used in circuit quantum electrodynamics (QED) to study light-matter interaction,\,\cite{Schoelkopf_2008,Niemczyk_2010} as quantum bus\,\cite{Kubo_2011,Mariantoni_2011} or photon storage devices.\,\cite{Hofheinz_2009} For many of those applications, the coherence time limited by internal loss channels of the resonator should be as long as possible. In the single photon limit, two-level states (TLSs) located at the metal/substrate, metal/air, and substrate/air interfaces are considered to be the main contributors to microwave losses.\,\cite{Wisbey_2010,Wenner_2011} In order to reduce these losses, much effort has been put into material development focusing on different metallization compounds\,\cite{Sage_2011,Ohya_2014} or substrate materials.\,\cite{OConnell_2008,Arzeo_2014} Also, quasiparticle generation in the superconducting material from stray infrared light\,\cite{Barends_2011} or thermal activation\,\cite{Gao_2008} generates losses in these resonators. In addition to material choice and screening quality, the sample design itself can have a large influence on the internal quality factor of the resonator.\,\cite{Lindstrom_2009,Khalil_2012} This influence can arise from impedance mismatches at the coupling ports resulting in Fano resonances,\,\cite{Fano_1961,Megrant_2012} from parasitic modes in the substrate,\,\cite{Wenner_2011a} or from addtitional resistive loss channels.\,\cite{Hornibrook_2012} For efficiently shielded and well designed setups based on optimized materials, internal quality factors above one million have been reached.\,\cite{Megrant_2012}\\
From a material perspective, Nb and Al are the workhorse materials in circuit QED experiments, which makes it important to probe and quantify the different loss channels in these materials and combinations thereof. In circuit QED experiments, often Al based Josephson junctions are integrated into Nb based coplanar waveguide (CPW) resonators.\,\,\cite{Niemczyk_2010,Baust_2015,Shcherbakova_2015} In these circuits, the junctions which are galvanically coupled to the resonator can contribute to microwave losses.\,\cite{Watanabe_2009} However, also the Nb/Al interface can be a possible loss channel. To evaluate this effect, we analyze these interfaces in detail and find that they can be a dominant source for TLS losses and -- above \SI{200}{\milli\kelvin}-- also for quasiparticle losses. In addition, we find the well known losses due to TLSs in CPW resonators fabricated on a Si/SiO$_2$ substrate. For these TLSs, we observe a temperature dependence in the characteristic saturation power of the TLSs in agreement with TLS theory.\,\cite{Leggett_1987,Phillips_1987} Galvanically coupled Josephson circuits are typically controlled via an external magnetic field, which can be a source for undesirable flux lines in large superconducting structures.\,\cite{Nsanzineza_2014} Additionally, superconducting enclosures often interfere with the need of applying magnetic control fields. To solve these issues, one can use a normal conducting layer on the sample backside. To benchmark this solution, we analyze losses introduced by eddy currents in a conductive (silver glue) material on the sample backside. Our analysis provides additional flexibility in the choice of sample package and fabrication process because we show that eddy currents can be avoided easily without using further superconducting materials by using thick enough substrates.
\begin{table*}
\caption{\label{tab:tab1}Overview of the samples analyzed in this work. The values of $\delta^{0}_{\mathrm{TLS}}, \delta_\text{c}, \beta$ and $P_\text{c}$ are obtained by fitting Eq.\,(\ref{eqn:Qi}) to a power sweep of each individual sample as shown in Fig.\,\ref{fig:07_Al-Bridge_CPW_vs_MS}. We also list confidence intervals generated by the fits. We obtain $\alpha$ by fitting Eq.\,(\ref{eqn:Pc}) to the values of $P_\text{c}(T)$. Sample~IX (MS) has a superconducting groundplane on the backside of the substrate.}
\includegraphics{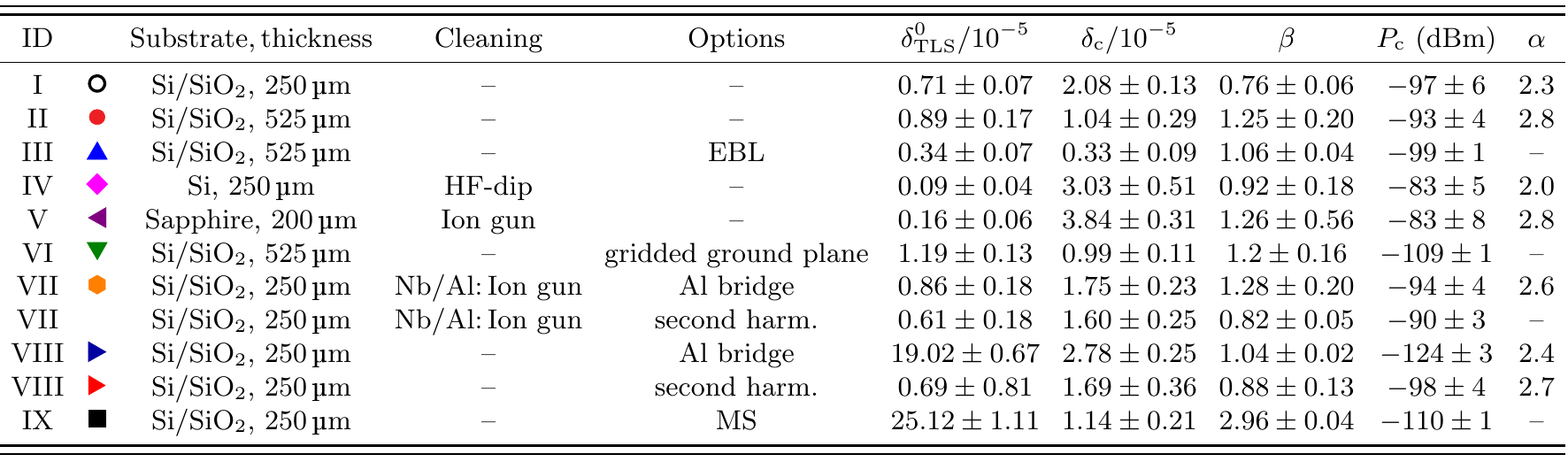}
\end{table*}
\section{Experimental techniques}
\begin{figure}[b]
\includegraphics{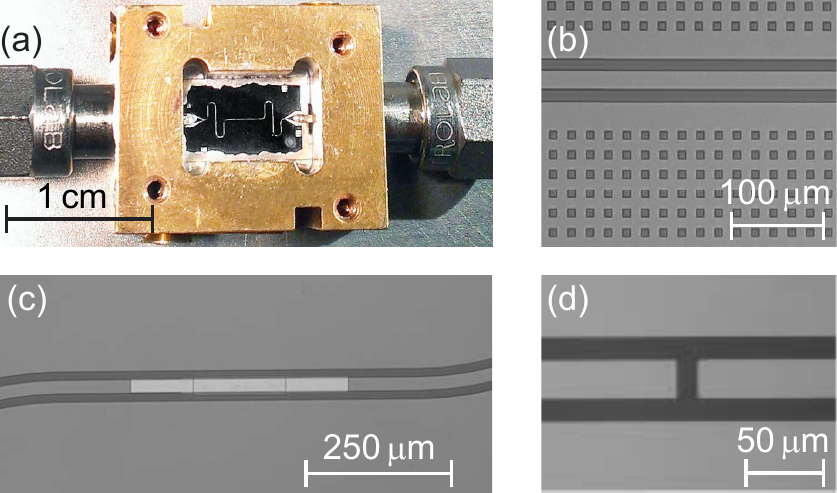}
\caption{\label{fig:07_Al-Bridge_CPW_vs_MS_sample}(a) Photograph of a resonator sample fixed with silver glue in a gold-plated copper box. (b) Micrograph of the gridded groundplane used for sample~VI. (c) Micrograph of sample~VII at the position where the Nb center conductor is replaced by an Al strip. (d) Micrograph of a typical coupling capacitor used for the CPW samples.}
\end{figure}

We first introduce our measurement setup and the most relevant fabrication steps. We systematically study the internal losses of one microstrip (MS) and eight CPW resonators with parameters as summarized in Table\,\ref{tab:tab1}. For electrical characterization, all samples discussed here are fixed with silver glue inside a gold-plated copper box as shown in Fig.\,\ref{fig:07_Al-Bridge_CPW_vs_MS_sample}\,(a) and cooled down to the temperature of the cold-stage of a dilution refrigerator which can be stabilized between \SI{50}{\milli\kelvin} and \SI{600}{\milli\kelvin} with a precision of $\pm$\SI{0.1}{\milli\kelvin}. Our low temperature setup has a radiation shield at \SI{700}{\milli\kelvin}, a cryoperm shield at \SI{4.2}{\kelvin}, and ${\mathrm{\mu}}$-metal shielding at room temperature. The input lines are heavily attenuated at different temperature levels in order to achieve populations of less than one photon on average inside the resonator. In the output lines we use cryogenic circulators mounted on the sample stage and on the \SI{700}{\milli\kelvin} plate for this purpose. We amplify the signal using a high electron mobility transistor amplifier at \SI{4.2}{K} and one additional room temperature amplifier. Measurements are performed with a vector network analyzer (VNA).\\
We investigate half-wavelength resonators which have resonance frequencies of approximately \SI{4}{\giga\hertz}. Our samples are fabricated either on a SiO$_2$-covered silicon, HF treated Si/SiO$_2$, or a sapphire substrate.\,\cite{Niemczyk_2009} The SiO$_2$ is thermally grown and the silicon is undoped with a specific resistance larger than \SI{1}{\kilo\ohm\centi\meter}.
 Onto these substrates we sputter-deposit a \SI{100}{\nano\meter}~thick Nb film. The resonators are patterned into these films using optical lithography (OL) or electron beam lithography (EBL) and reactive ion etching. The CPW resonators are comprised of a $s\,{=}\,\SI{20}{\micro\meter}$ wide center conductor separated from the ground plane by a $w\,{=}\,$\SI{12}{\micro\meter}~wide gap on either side. The MS resonator has a conductor width of \SI{200}{\micro\meter} and a \SI{100}{\nano\meter}~thick Nb layer on the backside of the substrate serving as ground. We compare samples with different substrate thicknesses and surface treatments to our standard process for Nb on a \SI{250}{\micro\meter}~thick silicon substrate covered with \SI{50}{\nano\meter} SiO$_2$ on both sides (cf.~sample~I in Tab.\,\ref{tab:tab1}). For sample~II and sample~III, we use  a substrate which is \SI{525}{\micro\meter}~thick and we compare OL and EBL. As a surface treatment, we remove the SiO$_2$ layer on top of the silicon substrate for sample~IV using hydrofluric acid (HF). Sample~V is fabricated on a \SI{200}{\micro\meter} thick sapphire substrate, which is in-situ cleaned by Ar ion beam etching before Nb sputter deposition. Sample~VI has a gridded groundplane as shown in Fig.\,\ref{fig:07_Al-Bridge_CPW_vs_MS_sample}\,(b). Sample~VII and sample~VIII are used to investigate the influence of Nb/Al interfaces in the center conductor of the resonator. For this purpose, we replace a \SI{150}{\micro\meter}~long piece of the Nb center conductor by an Al strip of identical width and thickness as shown in Fig.\,\ref{fig:07_Al-Bridge_CPW_vs_MS_sample}\,(c). The Al strip shares an \SI{100}{\micro\meter} long overlap with the Nb center conductor and is evaporated in an extra fabrication step. For sample~VII, we additionally clean the Nb surface by means of in-situ Ar-ion milling before the Al evaporation to remove oxides and resist residues from the Nb surface. More details on the fabrication process are given in appendix\,\ref{sec:fab}.
\section{Methods}
\begin{figure}[t]
\includegraphics[]{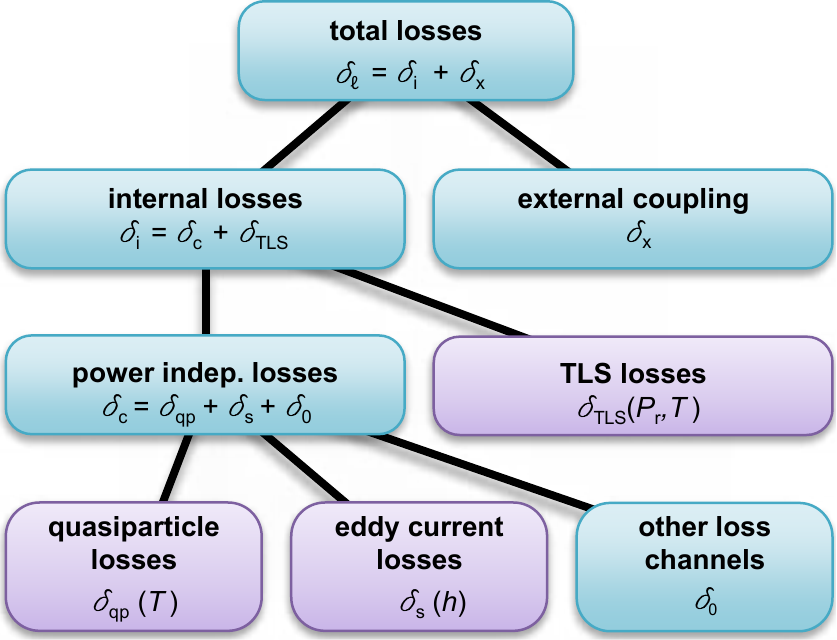}
\caption{\label{fig:02_qcontribs}Microwave loss contributions. Due to our knowledge of the contribution $\delta_\mathrm{x}$ from the coupling capacitors, our measurements of the total loss allows for a detailed discussion of TLS, quasiparticle, and eddy current losses (violet boxes). The quasiparticle losses are obtained from the Nb/Al samples and the eddy current losses via their dependence on the substrate thickness $h$.}
\end{figure}
\begin{figure}[t]
\includegraphics{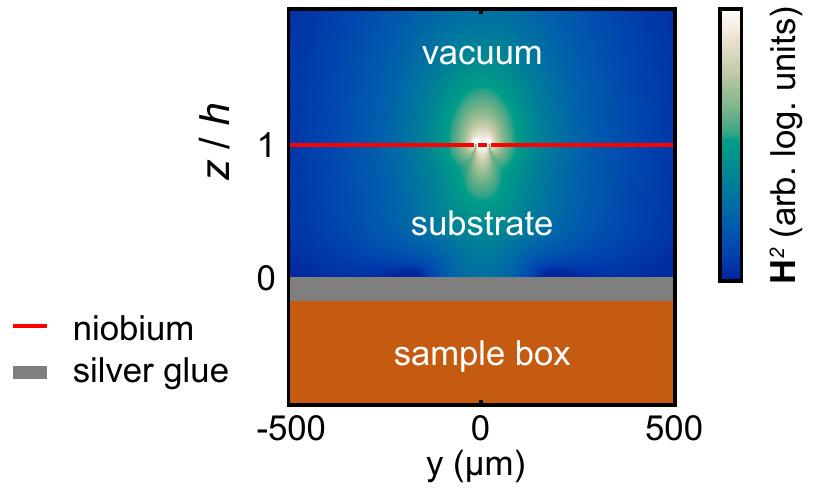}
\caption{\label{fig:05}Energy density (color coded) associated with the resonator magnetic field $\mathbf{H}(y,z)$ along a $yz$-cut through a CPW sample. Depicted in red is the Nb resonator at $z/h\,{=}\,1$ with its center conductor at $y\,{=}\,\SI{0}{\micro\meter}$. Underneath the sample we use silver glue to attach the substrate to the sample box.}
\end{figure}

In this section we discuss the different contributions to microwave losses which we take into account in our analysis. An overview of these contributions is given in Fig.\,\ref{fig:02_qcontribs}.  In order to compare the internal losses of devices with different resonance frequencies $\omega_{\mathrm{r}}/2\pi$, we calculate the internal quality factor as $Q_\text{i}\,{=}\,1/\left(Q_\mathrm{\ell}^{-1}{-}Q_\text{x}^{-1}\right)$. Here, $Q_\mathrm{\ell}$ and $Q_\mathrm{x}$ are the loaded and the external quality factor, respectively. From VNA spectroscopy measurements we obtain the loaded quality factor $Q_\mathrm{\ell}\,{=}\,\omega_{\mathrm{r}}/\Delta\omega$, where $\Delta\omega$ is the full width at half maximum of the Lorentzian peak. We use a combination of microwave simulations and reference measurements to obtain $Q_\text{x}\,{\simeq}\,\num{3e5}$ for the CPW samples. This external quality factor applies for all resonators as they share the same layout for the coupling capacitor shown in Fig.\,\ref{fig:07_Al-Bridge_CPW_vs_MS_sample}\,(d).\\
For our analysis, we calculate the losses $\delta_\text{i}\,{=}\,1/Q_\text{i}$, where $\tan\delta_\text{i}$ is the well-known loss tangent. We split the internal losses into a power independent term $\delta_{\mathrm{c}}$ and a power and temperature dependent TLS contribution $\delta_{\mathrm{TLS}}(P_{\mathrm{r}},T)$. Hence, the internal losses are described as
\begin{equation}
\delta_{\mathrm{i}}(P_{\mathrm{r}},T)=\delta_{\mathrm{TLS}}(P_{\mathrm{r}},T)+\delta_{\mathrm{c}}\,.
\label{eqn:Qi}
\end{equation}
Here, $P_{\mathrm{r}}\,{=}\,PQ_{\mathrm{\ell}}^2/n\pi Q_\text{x}$ is the power circulating inside the resonator\,\cite{Sage_2011} for the $n^{\mathrm{th}}$ mode and $P$ is the power resonantly applied to the input of the resonator. In Eq.\,(\ref{eqn:Qi}) $\delta_{\mathrm{TLS}}(P_{\mathrm{r}},T)$ describes losses due to the coupling of an ensemble of microscopic TLSs to the electromagnetic field of the resonator. These TLSs are located inside the dielectric and at the dielectric/metal interfaces in the vicinity of the resonator. For such a system, the TLS contribution reads\,\cite{Strom_1978,Sage_2011}
\begin{equation}
 \delta_{\mathrm{TLS}}(P_{\mathrm{r}},T)=\delta_{\mathrm{TLS}}^{0}\frac{\tanh\left(\hbar\omega_{\mathrm{r}}/2k_{\mathrm{B}}T\right)}{\sqrt{1+\left(P_{\mathrm{r}}/P_{\mathrm{c}}\right)^{\beta/2}}}\,.
\label{eqn:QTLS}
\end{equation}
In this expression, the exponent $\beta$ is known to be design-dependent,\,\cite{Wang_2009} $P_\text{c}$ is a characteristic power depending on the TLS properties,\,\cite{Schickfus_1977,Arzeo_2014} and $k_{\mathrm{B}}$ is the Boltzmann constant. Due to the distributed nature of the CPW resonator, TLS saturation does not occur uniformly across the sample but starts at voltage antinodes of the driven mode. Additionally, it depends on the center conductor width $s$ as well as on its separation to the groundplane $w$, which determine the electric field strength.\,\cite{Gao_2008_1,Sage_2011} In the low temperature and low power limit, the internal losses approach $\delta_{\mathrm{TLS}}^{0}$ which is mainly limited by the unsaturated TLSs. The characteristic power $P_\text{c}$ is proportional to $(\tau_{1}\tau_{\phi})^{-1}$, where $\tau_{1}$ and $\tau_{\phi}$ represent the relaxation time of the TLSs and the broadening of the levels due to their mutual interaction, respectively.\,\cite{Schickfus_1977} In the spin-boson model, the temperature dependence of $\tau_{1}^{-1}$ follows a $\coth(\hbar\omega/2k_{\mathrm{B}}T)$ dependence\,\cite{Jaeckle_1972,Leggett_1987} which is proportional to $T$ for $k_{\mathrm{B}}T\,{\gg}\,\hbar\omega$. Due to phonon mediated interaction between the TLSs one expects $\tau_{\phi}^{-1}\,{\propto}\,\tau_{\mathrm{ac}}^{-1}T\,{+}\,\tau_{\mathrm{op}}^{-1}/[\exp(\hbar\omega/k_{\mathrm{B}}T){-}1]$, where $\tau_{\mathrm{ac}}^{-1}$ and $\tau_{\mathrm{op}}^{-1}$ describe the TLS coupling rate to acoustic and optical phonons, respectively.\,\cite{Rudin_1990,Borri_1999} For low temperatures, $k_{\mathrm{B}}T\,{\ll}\,\hbar\omega$, the interaction is predominantly mediated by the term accounting for acoustic phonons $\tau_{\mathrm{ac}}^{-1}T$. However, in the regime $\hbar\omega\,{\simeq}\,k_{\mathrm{B}}T$ which is relevant for our experiments, a power law ${\propto}\,T^{\alpha}$ has been found for both, $\tau_{1}^{-1}$ and $\tau_{\phi}^{-1}$.\,\cite{Jaeckle_1972,Strom_1978,Schickfus_1977,Phillips_1987,Borri_1999,Rudin_1990,Lisenfeld_2007,Lisenfeld_2010} Therefore, the temperature dependence of the characteristic power can be approximated as\,\cite{Schickfus_1977}
\begin{equation}
 P_{\mathrm{c}}(T)=\frac{3\hbar^{2}\varepsilon}{2d^{2}\tau_{1}(T{=}0)\tau_{\mathrm{ac}}}\coth\left(\frac{\hbar\omega_{\mathrm{r}}}{2k_{\mathrm{B}}T}\right)T^\alpha\,,
\label{eqn:Pc}
\end{equation}
where $\varepsilon\,{=}\,\varepsilon_{0}\varepsilon_{\mathrm{r}}$ is the absolute permittivity of the dielectric and $d$ is the effective dipole moment of the TLSs. For resonators made from a single metal layer, these TLSs couple mainly to the electric field $\mathbf{E}_{\mathrm{r}}$ generated by the CPW structure. This situation changes for resonators including metal/metal interfaces, e.g.~Nb/Al interfaces, especially if these interfaces include an oxide layer of finite thickness $h_{\mathrm{ox}}$. In this case, the interface forms a Josephson junction and the TLSs also couple to the electric field $|\mathbf{E}_{\mathrm{Nb/Al}}|\,{=}\,V_{\mathrm{J}}/h_{\mathrm{ox}}$ inside this junction.\,\cite{Martinis_2005} Here, $V_{\mathrm{J}}\,{=}\,L_{\mathrm{J}}\partial I/\partial t$ is the voltage drop across the junction induced by the resonator current $I\,{=}\,I_{\omega_{\mathrm{r}}}\cos(\omega_{\mathrm{r}}t)$. The Josephson inductance is given by $L_{\mathrm{J}}\,{=}\,\hbar/2eI_{\mathrm{c}}\cos\varphi$, where $e$ is the electron charge, $I_{\mathrm{c}}$ the junction critical current and $\varphi$ the phase drop across the junction. Hence, the root mean square electric field across the junction reads as
\begin{equation}
 E_{\mathrm{Nb/Al}}^{\mathrm{rms}}=\frac{\hbar\omega_{\mathrm{r}}}{2eh_{\mathrm{ox}}\cos\varphi}\frac{I_{\omega_{\mathrm{r}}}}{\sqrt{2}I_{\mathrm{c}}}\,.\label{eqn:Vj}
\end{equation}
Equation (\ref{eqn:Vj}) shows that a finite resonator current amplitude $I_{\omega_{\mathrm{r}}}$ leads to an electric field across the junction which couples to possible TLSs. Therefore, we can apply Eq.\,(\ref{eqn:QTLS}) to describe the losses generated by TLSs inside a Josephson junction which is formed by an oxidized metal/metal interface incorporated into the center conductor of a CPW resonator.\\
In the following, we focus on the high power regime where the internal losses are limited by
\begin{equation}
\delta_{\mathrm{c}}(T,h)=\delta_{\mathrm{qp}}(T)+\delta_{\mathrm{s}}(h)+\delta_{0}\,.
\label{eqn:Qqp}
\end{equation}
Here, $\delta_{\mathrm{qp}}(T)$ are thermally induced quasiparticle losses and $\delta_{\mathrm{s}}(h)$ are eddy current losses on the backside of the substrate. The term $\delta_{0}$ comprises all other loss processes such as radiation,\,\cite{Kasilingam_1983} the finite surface resistance of superconductors,\,\cite{Noguchi_2012} and nonthermal quasiparticles generated by stray infrared light.\,\cite{Barends_2011,Gao_2008} Whereas dielectric losses dominate at very low temperatures, the losses related to the superconducting material typically become dominant when the sample temperature exceeds approximately $10\,\%$ of the critical temperature $T_{\mathrm{c}}$ of the superconductor.\,\cite{Arzeo_2014} In the temperature range of our experiments, we expect a considerable quasiparticle contribution only for samples containing Al. Assuming that the superconducting material is in the dirty or local limit, the quasiparticle contribution can be described with the Matthis-Bardeen theory,\,\cite{Mattis_1958,Gao_2008}
\begin{equation}
 \delta_{\mathrm{qp}}(T)=\frac{2\gamma}{\pi}\frac{e^{-\zeta}\sinh(\xi)K_{0}(\xi)}{1-e^{-\zeta}\left(\sqrt{2\pi/\zeta}-2e^{-\xi}I_{0}(\xi)\right)}\,.
\label{eqn:dQ}
\end{equation}
Here, $\gamma$ is the ratio of kinetic inductance to total inductance of the conductor, $\zeta\,{=}\,\Delta_{0}/k_{\mathrm{B}}T$ with the superconducting energy gap $\Delta_{0}$, $\xi\,{=}\,\hbar\omega_{\mathrm{r}}/2k_{\mathrm{B}}T$, and $I_{0}\,{,}\,K_{0}$ are the modified Bessel function of the first and second kind, respectively.\\
Eddy current losses $\delta_{\mathrm{s}}(h)$ in Eq.\,(\ref{eqn:Qqp}) mainly arise from the finite conductivity of the material used to fix our samples in the sample box. The thickness dependence results from a residual magnetic field $\mathbf{H}_{0}\,{\equiv}\,\mathbf{H}(z{=}0)$ on the backside of the substrate. Therefore, we also find a finite field $\mathbf{H}_{\mathrm{s}}$ in the volume of the silver glue used to fix our samples in the sample box (see Fig.\,\ref{fig:05}). Inside the silver glue volume, the field decays exponentially in $z$-direction for microwave frequencies within the skin depth $\lambda$, thus $\mathbf{H}_{\mathrm{s}}\,{=}\,\mathbf{H}_{0}\exp(-|z|/\lambda)$. The dissipated power due to eddy currents reads\,\cite{Morgan_1949} $P_\mathrm{s}\,{=}\,(1/2\sigma_{\omega_{\mathrm{r}}})\int(\nabla{\times}\mathbf{H}_{\mathrm{s}})^{2}\mathrm{d}V$, where $\sigma_{\omega_{\mathrm{r}}}$ is the electrical conductivity at the resonator frequency and $V$ is the volume of the conductive material on the backside of the substrate. Even though $P_\mathrm{s}$ would diverge for insulators, it is vaild for the conductive materials used in our work. To evaluate the thickness dependence, we calculate the field components of the magnetic field $\mathbf{H}(x,y,z)$ in the substrate following Ref.\,\onlinecite{Simons_1982}. We modify this model by appropriate boundary conditions for the conductor-backed substrate demanding the $z$-component of $\mathbf{H}_{0}$ to vanish. The $x$-component of $\mathbf{H}_{0}$, which points in the direction along the transmission line, can also be neglected in comparison with the $y$-component. Using these assumptions, the field on the backside of the substrate can be written as $\mathbf{H}_{0}(x,y)\,{\approx}\,\hat{e}_{\mathrm{y}}|\mathbf{H}_{0}(x,y)|$. Assuming a sinusodial current distribution along the transmission line, we find $|\mathbf{H}_{0}(x,y)|\,{=}\,H_{0}k(y,h)\sin\left(n\pi x/L\right)$, where $L$ is the resonator length and $k(y,h)$ is the field distribution at $z\,{=}\,0$ along the $y$-direction. The maximum field strength is given as $H_{0}\,{=}\,\sqrt{(1+\varepsilon_{\mathrm{r}})P_{\mathrm{r}}Z_{0}/2}(Z_{\mathrm{vac}}w)^{-1}$. Here, $Z_{0}\,{=}\,\SI{50}{\ohm}$ and $Z_{\mathrm{vac}}\,{\simeq}\,\SI{377}{\ohm}$ are line and vacuum impedance, respectively, and $\varepsilon_{\mathrm{r}}\,{=}\,11.9$ is the relative dielectric constant of the Si substrate. Finally, we calculate the dissipated power
\begin{align}
P_\mathrm{s}(h)     &= \int\frac{\mathrm{d}V}{2\sigma_{\omega_{\mathrm{r}}}}\left(\nabla\times\hat{e}_{\mathrm{y}}\frac{H_{0}k(y,h)\sin\left(n\pi x/L\right)}{\exp(|z|/\lambda)}\right)^2 \notag\\
    &\approx \sqrt{\frac{\omega_{\mathrm{r}}\mu}{2\sigma_{\omega_{\mathrm{r}}}}}\frac{(1+\varepsilon_{\mathrm{r}})P_{\mathrm{r}}Z_{0}L}{32(Z_{\mathrm{vac}}w)^{2}}K(h)^{2}\,.
\label{eqn:Ph}
\end{align}
In the above expression, $\int\mathrm{d}V\,{=}\,\int_{0}^{L}\mathrm{d}x\int_{-\infty}^{\infty}\mathrm{d}y\int_{0}^{-\infty}\mathrm{d}z$, $\mu\,{\simeq}\,\mu_{0}$ is the absolute permeability of the conductive material,\,\cite{Pelco_2013} $K(h)^2\,{=}\,\int_{-\infty}^{\infty}k(y,h)^2\mathrm{d}y$ and we use $\lambda\,{=}\,\sqrt{2/\omega_{\mathrm{r}}\mu\sigma_{\omega_{\mathrm{r}}}}$.\,\cite{Maxwell_1947} The thickness dependent losses $\delta_{\mathrm{s}}(h)$ are defined as $P_{\mathrm{s}}{/}P_\mathrm{r}$, the ratio of the power dissipated in the system to stored power. Hence,
\begin{equation}
\delta_{\mathrm{s}}(h)=\sqrt{\omega_{\mathrm{r}}\mu/2\sigma_{\omega_{\mathrm{r}}}}(1+\varepsilon_{\mathrm{r}})Z_{0}K(h)^2L/32(Z_{\mathrm{vac}}w)^{2}\,,\label{eqn:Q_s}
\end{equation}
which is independent of the power circulating inside the resonator.
\section{Results}
\begin{figure}[t]
\includegraphics[]{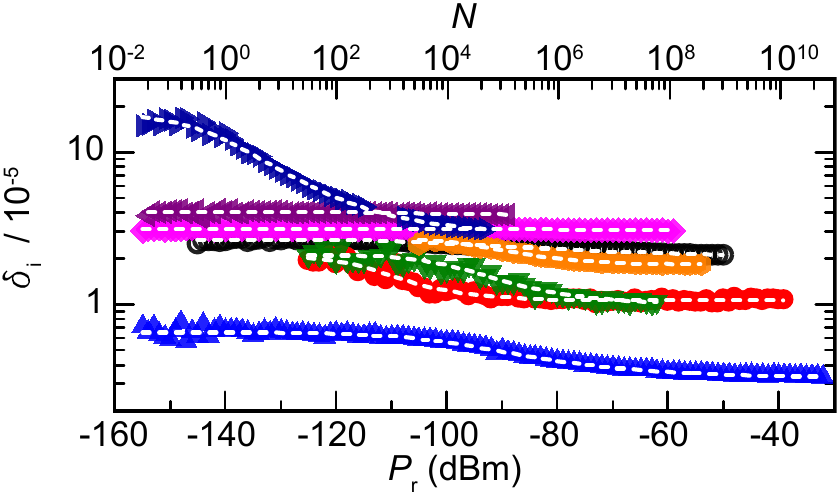}
\caption{\label{fig:07_Al-Bridge_CPW_vs_MS}Internal losses for samples~I\,--\,VIII plotted versus the microwave power circulating inside the resonator at \SI{50}{\milli\kelvin}. Symbols are explained in Tab.\,\ref{tab:tab1}. The average photon number on the top axis scale is calculated as $N\,{=}\,2\pi P_{\mathrm{r}}/\hbar\omega_{\mathrm{r}}^2$. Dashed lines are fits of Eq.\,(\ref{eqn:Qi}) to the data.}
\end{figure}
In the following, we present the experimental results starting with a comparison of TLS related losses. Next, we show how losses are affected by Nb/Al interfaces which are necessary for the galvanic coupling of Al-based Josephson circuits to the Nb center conductor of the CPW. Finally, we analyze the influence of a finite substrate thickness on internal losses.
\subsection{TLS losses in Nb resonators}
\label{sec:TLS}

\begin{figure}[t]
\includegraphics[]{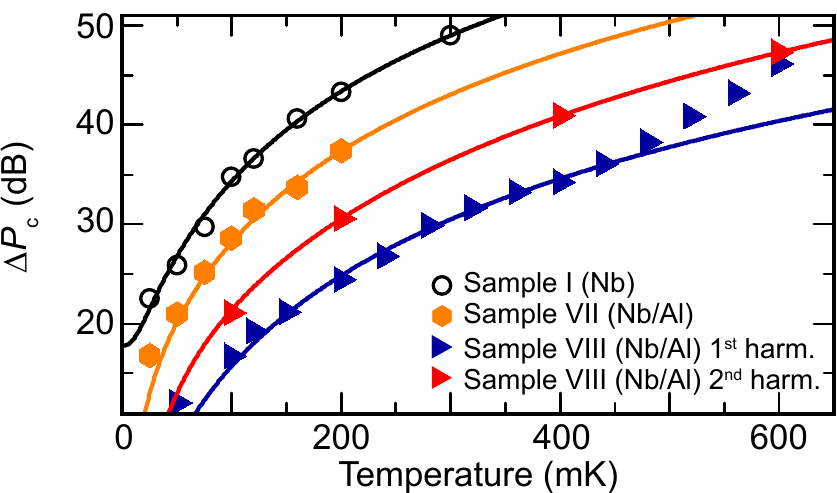}
\caption{\label{fig:05_Pc}Relative change $\Delta P_{\mathrm{c}}$ of the characteristic power} plotted versus temperature for different samples. Solid lines are fits of Eq.\,(\ref{eqn:Pc}) to the data. For the first harmonic mode of sample~VIII we observe a strong increase for temperatures above \SI{440}{\milli\kelvin}. For this dataset we fit only to datapoints below \SI{440}{\milli\kelvin}. For better visibility, there is an offset of \SI{6}{\deci\bel} between each dataset.
\end{figure}

In this subsection, we discuss the TLS losses for samples I\,--\,VI (pure Nb resonators). Fitting Eq.\,(\ref{eqn:Qi}) to the power dependence of  $\delta_{\mathrm{i}}$, we extract $\delta^{0}_{\mathrm{TLS}}$, $\beta$ and $P_{\mathrm{c}}$ (see Fig.\,\ref{fig:07_Al-Bridge_CPW_vs_MS}). The results are summarized in Table\,\ref{tab:tab1}. To obtain further insight into the nature of the TLSs, we analyze the relative change $\Delta P_{\mathrm{c}}\,{=}\,P_{\mathrm{c}}(T)\,{-}\,P_{\mathrm{c}}(0)$, which is linked to TLS properties via Eq.\,(3). For each individual sample, we observe an increase in $P_{\mathrm{c}}$ for increasing temperature due to the temperature dependent lifetime of the TLSs as shown in Fig.\,\ref{fig:05_Pc}. From a fit based on Eq.\,(\ref{eqn:Pc}), we find $\alpha\,{\simeq}\,2.5\pm0.3$ (average over samples~I\,--\,VIII). This value deviates from $\alpha\,{=}\,1$ expected from the spin-boson model,\,\cite{Jaeckle_1972,Leggett_1987} but is comparable to values reported for TLSs in glasses\,\cite{Phillips_1987,Schickfus_1977} and phase qubits.\,\cite{Lisenfeld_2007,Lisenfeld_2010} The deviation can be attributed to the fact that our experiments are not in the low temperature limit, but rather in the intermediate regime $\hbar\omega\,{\simeq}\,k_{\mathrm{B}}T$.\\
Next, we discuss the low temperature and low power losses $\delta_\text{TLS}^0$. For sample~I, which serves as a reference sample, we find a TLS contribution $\delta^{0}_{\mathrm{TLS}}\,{\simeq}\,\num{7.1e-6}$. We now study the influence of TLSs near the surface using sample~IV, where we have cleaned the surface with an HF-dip before metallization. For this sample, we measure $\delta^{0}_{\mathrm{TLS}}\,{\simeq}\,\num{9e-7}$, which is one order of magnitude lower than for sample~I. Hence, we conclude that most of the TLS losses are introduced by the bulk SiO$_2$ layer, the SiO$_2$/metal, and SiO$_2$/air surfaces. In contrast, losses at the metal/air interface are significantly smaller (${\le}\,\num{9e-7}$). For sample~VI (gridded groundplane) the TLS losses are comparable to those of a resonator with continuous groundplane.
\subsection{Losses due to TLSs at Nb/Al interfaces}
\label{sec:Losses}
\begin{figure}[t]
\includegraphics{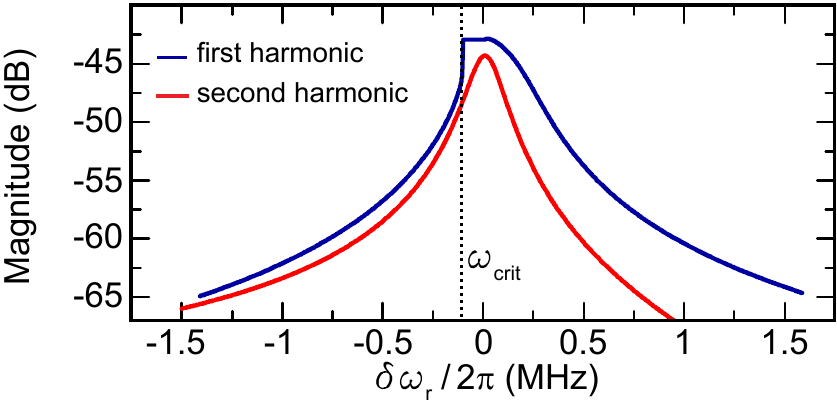}
\caption{\label{fig:06_nonlin}Transmission magnitude plotted versus $\delta\omega_{\mathrm{r}}\,{=}\,\omega-\omega_{\mathrm{r}}$ for the first and second harmonic mode of sample~VIII at $P_{\mathrm{r}}\,{\simeq}\,\SI{-45}{\deci\bel}$m. For the first harmonic mode, the resonance shows a nonlinear behavior due to the presence of the Nb/Al interface forming a Josephson junction. We use the characteristic frequency $\omega_{\mathrm{crit}}$ to determine the junctions critical current. The second harmonic mode does not couple to the the junction and therefore keeps its Lorentzian peak form.}
\end{figure}

\begin{figure}[t]
\includegraphics{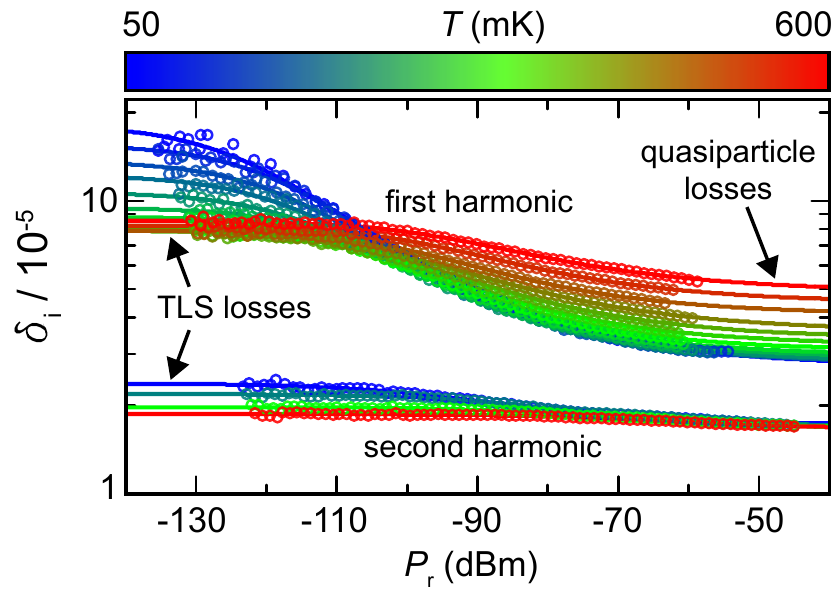}
\caption{\label{fig:07_Al-Bridge}Internal losses plotted versus the circulating microwave power for the first two harmonic modes of sample~VIII (no Ar ion cleaning) at temperatures between \SI{50}{\milli\kelvin} and \SI{600}{\milli\kelvin}. Solid lines are fits of Eq.\,(\ref{eqn:Qi}) to the data. Arrows indicate the regions relevant for the analysis of TLS and quasiparticle losses.}
\end{figure}

We use sample~VIII, where an Al strip is placed at the current antinode of the fundamental mode, to study the influence of Nb/Al interfaces introduced into the center conductor of CPW resonators. These interfaces are known to introduce losses in galvanically coupled Josephson junction based circuits.\,\cite{Watanabe_2009,Shcherbakova_2015} However, the TLS effect on the internal quality factor of superconducting resonators has not yet been quantified. The transmission of the first harmonic mode of the resonator without ion gun treatment shows a deviation from the Lorentzian lineshape for large probe power (see Fig.\,\ref{fig:06_nonlin}). As this behavior is typical for resonators including Josephson junctions,\,\cite{Manucharyan_2007,Watanabe_2009} we treat the Nb/Al interfaces as large-area Josephson junctions. As expected, the second harmonic mode which has a current node at the interface position shows a Lorentzian behavior. Due to the presence of the interfaces, we also observe a non-equidistant mode spacing, which is again typical for resonators including Josephson junctions.\,\cite{Niemczyk_2010} Specifically, we find $\omega_{\mathrm{r},2}/\omega_{\mathrm{r},1}\,{\simeq}\,1.93$, where $\omega_{\mathrm{r},n}$ is the resonance frequency of the $n^{\mathrm{th}}$ mode. In contrast, all samples without Nb/Al interface (samples~I--VI) show an equidistant mode spacing, $\omega_{\mathrm{r},n}/\omega_{\mathrm{r},1}\,{=}\,n$.\\
The presence of the oxidized interfaces results in large TLS losses. For the first harmonic mode of sample~VIII, we observe an increase of more than one order of magnitude in $\delta_\text{TLS}^0$ compared to sample~I (pure Nb) or sample~VII (cleaned Nb/Al interface). Hence, we draw two important conclusions. First, without cleaning step, the Nb oxides present at the interfaces are strong TLS sources. Second, these TLSs can cause significant losses. This behavior is not immediately obvious, because the Nb/Al interfaces are placed at a voltage node of the first harmonic mode. Consequently, the TLSs associated with the Nb oxides at these interfaces are not expected to couple to the resonator electric field. Nevertheless, due to the electric field $\mathbf{E}_{\mathrm{Nb/Al}}$ between the Nb and the Al layer in the overlap area, we observe a pronounced power and temperature dependence of $\delta_{\mathrm{i}}$ (see Fig.\,\ref{fig:07_Al-Bridge}). Actually, according to Eq.\,(\ref{eqn:Vj}), $E_{\mathrm{Nb/Al}}^{\mathrm{rms}}$ is proportional to the resonator current and therefore maximum if the Nb/Al interface is placed at the current antinode (voltage node) of the resonator field.\\
In the following, we discuss why the TLSs in the interfaces are not yet saturated by $E_{\mathrm{Nb/Al}}^{\mathrm{rms}}$ for a probe power corresponding to the vacuum current $I_{\omega_{\mathrm{r},1}}\,{=}\,\sqrt{\hbar\omega_{\mathrm{r}}^{2}/2\pi^{2}Z_{0}}\,{\simeq}\,$\SI{21}{\nano\ampere}, which is necessary to observe the power dependence shown in Fig.\,\ref{fig:07_Al-Bridge}. From the critical points $\omega_{\mathrm{crit}}(P_{\mathrm{r}})$ as indicated in Fig.\,\ref{fig:06_nonlin}, we can derive the critical current\,\cite{Manucharyan_2007} $I_{\mathrm{c}}\,{=}\,\sqrt{3}(Z_{0}\Phi_{0}P_{\mathrm{r}})^{1/3}/[2\pi\delta_{\mathrm{\ell}}L_{\mathrm{r}}(8\omega_{\mathrm{crit}})^{2/3}]\,{\simeq}\,\SI{19.2}{\micro\ampere}$, where $L_{\mathrm{r}}\,{\simeq}\,\SI{6}{\nano\henry}$ is the resonator inductance. With a junction area of approximately \SI{2000}{\square\micro\meter}, we obtain a current density $J_{\mathrm{c}}\,{\simeq}\,\SI{0.96}{\ampere\per\square\centi\meter}$. This very low critical current density is expected, as the Nb was exposed to air for two days.\,\cite{Kleinsasser_1995} Nevertheless, we obtain $I_{\mathrm{c}}\,{\gg}\,I_{\omega_{\mathrm{r},1}}$ and can use the approximation $\cos\varphi\,{=}\,1$. Assuming an oxide thickness on the order of \SI{1}{\nano\meter}, we obtain $E_{\mathrm{Nb/Al}}^{\mathrm{rms}}\,{\simeq}\,$\SI{6}{\volt\per\meter} from Eq.\,(\ref{eqn:Vj}). On the one hand, this field is large enough to mediate a coupling between the TLSs and the resonator. On the other hand, it is small compared to the experimentally observed saturation field $|\mathbf{E}_{\mathrm{Nb/Al}}^{\mathrm{c}}|\,{=}\,\SI{44}{\volt\per\meter}$ given by the resonator current $I_{\omega_{\mathrm{r}}}\,{\simeq}\,\SI{140+-40}{\nano\ampere}$ for $P_{\mathrm{c}}\,{\simeq}\,\SI{-124}{\deci\bel}$m.\\
In the next step, we confirm our model of TLS losses in the interfaces by analyzing the second harmonic mode of sample~VIII, which has a voltage antinode at the interface position. As shown in Fig.\,\ref{fig:07_Al-Bridge}, this mode shows significantly less internal losses than the first harmonic mode. We measure $\delta^{0}_{\mathrm{TLS}}{=}\num{7e-6}$ and $\delta_\text{c}{=}\num{1.7e-5}$, that is, values comparable to those of the first harmonic mode of the pure Nb resonator (sample~I). This observation is unexpected for losses based on a model considering only uniformly distributed loss mechanisms.\,\cite{Goeppl_2008} Instead, the effect can be explained by TLSs localized in the Nb/Al interfaces. Since for the second harmonic mode the interfaces are placed at the current node, $I_{\omega_{\mathrm{r},2}}\,{\simeq}\,0$, the corresponding electric field $E_{\mathrm{Nb/Al}}^{\mathrm{rms}}$  is vanishingly small. Therefore, only the TLSs outside the interface, which couple to the resonator electric field, introduce losses comparable to those of sample~I (pure Nb). We note that also the power independent contribution $\delta_\text{c}$ is smaller for the second harmonic mode than for the first harmonic mode. This behavior indicates that the interfaces are not only strong sources for TLS losses but also for local resistive losses.\\
Finally, we show that the local losses induced by the interfaces can be significantly reduced by in-situ ion gun treatment of the Nb surface before Al evaporation. Consequently, $\delta_{\mathrm{TLS}}^{0}\,{\simeq}\,\num{9E-6}$ of the fundamental mode of sample~VII (ion gun treatment) is one order of magnitude larger than for sample~VIII (no ion gun treatment). As expected, we find that  $\delta_{\mathrm{TLS}}^{0}$ of sample~VII is similar to $\delta_{\mathrm{TLS}}^{0}$ of both the second harmonic mode of sample~VIII and the fundamental mode of sample~I (pure Nb).
\subsection{Temperature dependent losses in Nb/Al interfaces}
\begin{figure}[t]
\includegraphics{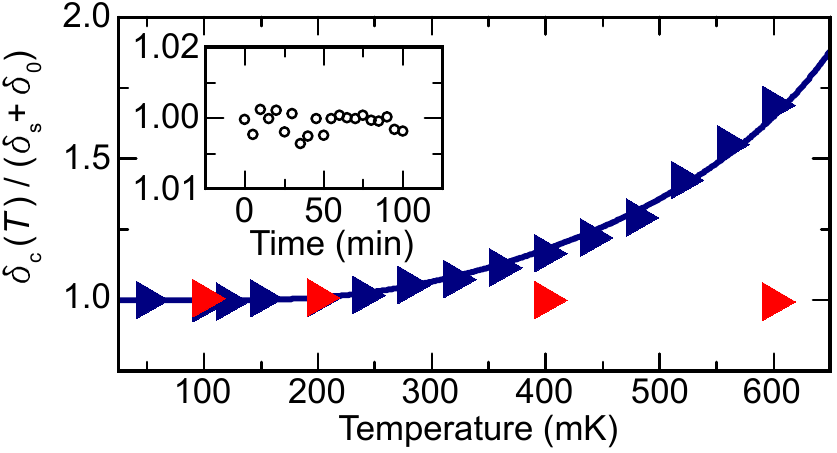}
\caption{\label{fig:08_tdepQc}Normalized losses $\delta_{\mathrm{c}}(T)/(\delta_{\mathrm{s}}\,{+}\,\delta_{0})$ of sample~VIII plotted versus sample temperature. Blue triangles correspond to the first and red triangles to the second harmonic mode. The solid line is a fit of $\delta_{\mathrm{c}}(T)$ using Eq.\,(\ref{eqn:Qqp}) based on calculations of $\delta_{\mathrm{qp}}(T)$. The inset shows the time dependence of $\delta_{\mathrm{c}}(T)/(\delta_{\mathrm{s}}\,{+}\,\delta_{0})$ at \SI{50}{\milli\kelvin}.}
\end{figure} 
We first analyze the temperature dependence $\Delta P_{\mathrm{c}}(T)$ for both samples including a Nb/Al interface (sample~VII and sample~VIII). As shown in Fig.\,\ref{fig:05_Pc}, the exponent $\alpha$ is in the same range as for all other samples (see Sec.\,\ref{sec:TLS}). Specifically, we extract $\alpha\,{\simeq}\,2.6$ (sample~VII), $\alpha\,{=}\,2.4$ (sample~VIII, first harmonic), and $\alpha\,{=}\,2.7$ (sample~VIII, second harmonic). When extracting $\alpha$ for the first harmonic of sample~VIII, we only take datapoints below \SI{440}{\milli\kelvin} into acount. We attribute the increased slope of $P_{\mathrm{c}}(T)$ above this temperature to self heating processes\,\cite{Lang_2003,Tholen_2007,Wang_2005,Guenon_2010} inside the interfaces. This effect is not present in the second harmonic of sample~VIII which has a current node at the interface position.\\
Next, we study the influence of the Nb/Al interfaces on $\delta_{\mathrm{c}}$, i.e., in the high power regime. For pure Nb resonators, we observe no significant change of $\delta_{\mathrm{c}}$ in the temperature range between \SI{50}{\milli\kelvin} and \SI{600}{\milli\kelvin}. This behavior is expected because the number of quasiparticles is negligible for our experiments due to the critical temperature $T_{\mathrm{c}}\,{\simeq}\,\SI{9}{\kelvin}$ of the Nb films. However, the situation is different for samples including an Al strip, which has a lower $T_{\mathrm{c}}\,{\simeq}\,\SI{1.5}{\kelvin}$. In Fig.\,\ref{fig:08_tdepQc}, we show the temperature dependence of $\delta_{\mathrm{c}}$ for the first two modes of sample~VIII. For the first harmonic mode, we observe a quasiparticle induced increase of $\delta_{\mathrm{c}}$, which becomes relevant for temperatures above \SI{200}{\milli\kelvin}. Using $\delta_{\mathrm{s}}\,{+}\,\delta_{0}\,{=}\,\num{3.6e4}$ obtained from a power sweep at \SI{50}{\milli\kelvin} where quasiparticles are negligible, we fit Eq.\,(\ref{eqn:Qqp}) to the data and find a kinetic inductance fraction of $\gamma\,{\simeq}\,\num{3.5e-4}$. This value is two orders of magnitude smaller than values reported in literature.\,\cite{Gao_2008} We explain this difference by the fact that the length of the Al strip is only $1/100$ of the total length of the center conductor. In contrast to the first harmonic mode, $\delta_\text{c}$ of the second harmonic mode shows no temperature dependence because the current distribution of the resonator has a node at the Al position in this case. Hence, quasiparticles in the Al do not carry a significant amount of the current circulating inside the resonator. Therefore, we conclude that also with respect to quasiparticle losses it is advantageous to place such interfaces at current nodes.\\
Due to the reduced superconducting energy gap of Al compared to Nb, nonequilibrium quasiparticles generated by stray infrared light can be trapped in the Al layer,\,\cite{Booth_1993,Court_2013} which reduces the number of quasiparticles in the Nb part of the center conductor. This effect can be used to decrease quasiparticle losses for resonator modes that have a current node at the Al layer. Indeed, we measure $\delta_{\mathrm{c}}\,{\simeq}\,\num{1.65E-5}$ for the second harmonic mode of both samples including Al layers, which is smaller than $\delta_{\mathrm{c}}\,{\simeq}\,\num{2.2E-5}$ measured for the second harmonic mode of sample~I (pure Nb resonator).
Due to variations in the experimental environment, loss rates can vary as a function of time at a minutes timescale.\,\cite{Burnett_2014,Muller_2015} In the inset of Fig.\,\ref{fig:08_tdepQc} we show the fluctuations of $\delta_{\mathrm{c}}$ over a $100$ minute time interval. From the statistics we evaluate a relatively small standard deviation \num{3.8E-3}. Also over a period of several days we do not observe any significant change in the resonator losses.
\subsection{Eddy current losses} 
\begin{figure}[t]
\includegraphics{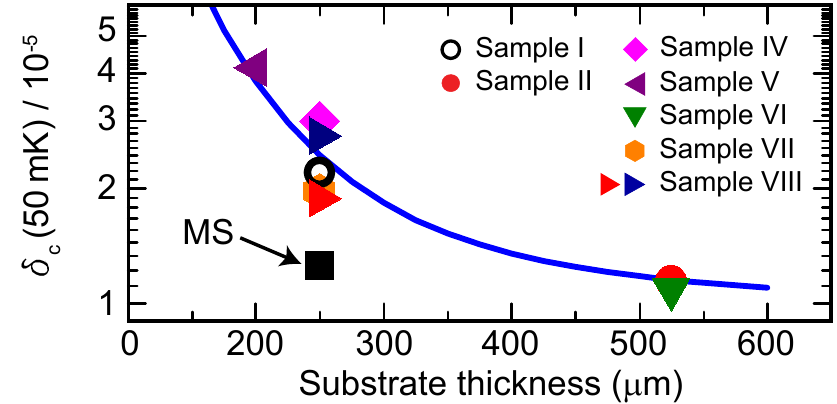}
\caption{\label{fig:09_thickness}Power independent losses $\delta_{\mathrm{c}}$ measured at \SI{50}{\milli\kelvin} plotted as a function of the substrate thickness. The solid line is a fit  based on Eq.\,(\ref{eqn:Qqp}) and Eq.\,(\ref{eqn:Q_s}) not accounting for sample~IX (MS), which has a superconducting layer at the sample backside.}
\end{figure}
In this subsection we analyze losses resulting from eddy currents which are induced in the conductive material on the backside of the substrate. For this analysis, we compare all CPW samples fabricated by optical lithography. We observe that $\delta_{\mathrm{c}}$ shows a significant dependence on the substrate thickness as displayed in Fig.\,\ref{fig:09_thickness}. Assuming a negligible quasiparticle contribution $\delta_{\mathrm{qp}}(\SI{50}{\milli\kelvin})\,{\simeq}\,0$, a numerical fit based on Eq.\,(\ref{eqn:Qqp}) and Eq.\,(\ref{eqn:Q_s}) yields $\delta_{0}\,{\simeq}\,\num{8e-6}$ and $\sigma_{\omega_{\mathrm{r}}}\,{\simeq}\,$\SI{7e7}{\siemens\per\meter}. This conductivity is approximately hundred times larger than the room temperature conductivity of our silver glue\,\cite{Pelco_2013} and comparable to the room temperature conductivity of copper.\,\cite{Maxwell_1947,Thorp_1954} Since our experiments are carried out at low temperatures where the conductivity of metals typically increases by a factor of $100$,\,\cite{Matula_1979,Fouaidy_2006} we can quantitatively explain the loss behavior shown in Fig.\,\ref{fig:09_thickness} by ohmic losses in the silver glue. The influence of eddy current losses depends on the material present underneath the sample. Compared to pure metals, silver glue has a relatively low conductivity and therefore larger losses. For samples with a substrate thickness of \SI{525}{\micro\meter} losses are already enhanced by $13\,\%$ compared to $\delta_{0}$ in the high power regime. Samples fabricated on \SI{200}{\micro\meter} thick substrates show a loss increase by a factor of four compared to $\delta_{0}$. The slight scatter in $\delta_{\mathrm{c}}$ for $h\,{=}\,$\SI{250}{\micro\meter} is attributed to our assumption of a universal $\delta_{0}$. This assumption is, of course, only a rough estimation because the samples are fabricated with different cleaning methods.\\
In summary, we can quantitatively explain losses due to eddy currents using numerical calculations of the $\mathbf{H}$-field distribution in conductor backed CPW structures. The electrodynamical model suggests that it is advantageous to either use thick substrates or materials with high conductivity on the backside of the substrate. Therefore, losses caused by eddy currents can be avoided by using superconducting materials at this position. This conclusion is supported by our measurements on the MS resonator (sample~IX), which is fabriacated on a \SI{250}{\micro\meter} thick substrate and employs a superconducting ground plane. For this sample, we measure $\delta_{\mathrm{c}}\,{\simeq}\,\num{1e-5}$, which is reduced compared to the values of the other samples fabricated on a \SI{250}{\micro\meter} thick substrate by a factor of two, despite the fact that the field at the bottom surface of the substrate is an order of magnitude larger compared to the CPW samples.\\

In contrast to other groups,\,\cite{Quintana_2014} we observe a decrease of the power independent losses $\delta_{\mathrm{c}}$ by a factor of three for the sample fabricated with EBL (sample~III) compared to sample~II, which is fabricated with OL. There are two likely reasons for this observation. First, the Nb edges are smoother for the EBL sample than for the OL samples, resulting in a reduced field elevation at these positions. Second, compared to our OL process, the PMMA resist used for EBL may leave less resist residuals on the sample. Such resist residuals are known to introduce losses.\,\cite{Quintana_2014}
\section{Conclusions}
We fabricate superconducting CPW resonators and analyze mechanisms leading to microwave losses. With respect to TLS losses, we find that the characteristic power necessary to saturate TLSs shows a pronounced temperature dependence. Specifically, we extract an exponent $\alpha\,{=}\,2.5$ for the polynomial part of this temperature dependence. Furthermore, we show that an HF-dip can be used to remove TLSs and therefore reduce losses on oxide covered silicon substrates. In addition, we investigate resonators containing Nb/Al interfaces, which typically occur for galvanically coupled quantum circuits. We quantitatively investigate the temperature and power dependence of the losses caused by these interfaces. These losses can be reduced by either placing the interfaces in a current node or by removing the oxide layer containing the TLSs by in-situ Ar ion milling. In the high power limit, where the TLSs are saturated, we observe temperature dependent losses induced by quasiparticle excitations in the Al. Finally, we investigate losses generated by eddy currents in the conductive material on the backside of the substrate. Our results show that the usage of thick enough substrates or metals with high conductivity on the substrate backside helps to minimize these losses.\\
\\
This work is supported by the German Research Foundation through SFB 631 and FE 1564/1-1, EU projects CCQED, PROMISCE, the doctorate program ExQM of the Elite Network of Bavaria.
\bibliography{Goetz_lossmechanisms_bib}

\appendix

\section{Sample fabrication}
\label{sec:fab}

Here, we give a detailed description of our fabrication processes. All substrates have lateral dimensions of $\SI{6}{\milli\meter}\,{\times}\,\SI{10}{\milli\meter}$ and are cleaned by a series of isopropanol and acetone dips in an ultrasonic bath. We sputter deposit the \SI{100}{\nano\meter} thick niobium films with a deposition rate of \SI{0.33}{\nano\metre\per\second} at an argon pressure of \SI{275}{\micro\bar} with an argon flow of $10$\,sccm.\\
For optical lithography we use AZ5214E resist, which is spin-coated onto the Nb with $1000$\,rpm. We perform mask exposure to ultraviolet light after baking for \SI{70}{\second} at \SI{110}{\celsius}. For electron beam lithography we use PMMA/MA\,33\,\% resist spin-coated with 2000 rpm for one minute. To activate the resist, we use a dose of \SI{200}{\micro\coulomb\per\square\centi\meter} and develop with AR 600-56 developer.\\
For reactive ion etching we use  an Ar/SF$_6$ plasma with a flow of 10\,sccm for Ar and 20 sccm for SF$_6$. We etch for \SI{70}{\second} at a pressure of \SI{20}{\milli\bar} using \SI{50}{\watt} for the ICP plasma.\\
For sample~IV, we remove the SiO$_2$ layer on top of the Si substrate. To this end, we wet-etch for three minutes with a ratio HF:H$_2$O\,=\,1:10. Afterwards, we rinse the sample with deionized water before placing it in the sputtering chamber within five minutes after the wet etch. Sample~V is additionally cleaned by in-situ Ar ion milling for \SI{60}{\second} before sputter deposition. Here, we use an acceleration voltage of \SI{100}{\volt}, an operating pressure of \SI{4E-6}{\milli\bar}, a filament current of \SI{2.7}{\ampere}, and an emission current of \SI{30}{\milli\ampere}. The square holes of sample~VI have a width of \SI{8}{\micro\meter} and are separated by \SI{12}{\micro\meter}. Aluminum evaporation for sample~VII and sample~VIII is done at \SI{2E-7}{\milli\bar} with \SI{8}{\kilo\volt} and a filament current of \SI{400}{\milli\ampere}, which results in a rate of \SI{12}{\angstrom\per\second}. To clean the Nb surface of sample~VII, we use an Ar ion milling process for \SI{60}{\second} with an Ar flow of $0.5$\,sccm, an emission current of \SI{20}{\milli\ampere}, an extraction voltage of \SI{600}{\volt}, and an acceleration voltage of \SI{2.4}{\kilo\volt}.

\end{document}